\documentstyle[referee]{laa}

\begin{document}

\thesaurus{10 
	   (10.11.1 
	   )}

\title{Kinematics of Galactic Disk}
\author{ Jozef Kla\v{c}ka }
\institute{Department of Astronomy and Astrophysics,
Faculty for Mathematics and Physics,
Comenius University,
Mlynsk\'{a} dolina,
842 15 Bratislava,
Slovak Republic, 
E-mail: klacka@fmph.uniba.sk}
\date{}
\maketitle
\begin{abstract}
The method of constructing of rotation curve for our galactic disk is
developed. Simple physical model shows that the velocity of the Sun around
the center of the Galaxy must be less than currently accepted IAU value
220 km/s. Independently obtained result for neutral hydrogen ($\lambda =$
21.1 cm) confirms the previous statement. The values of the Sun's velocity
is about 110 km/s. As a consequence, the rotation curve for the disk of our
Galaxy is not flat, but it is a decreasing function of the distance from the
center of the Galaxy. This conclusion seems to be consistent also with
other galaxies.
\end{abstract}

\section{Introduction}
One of the key arguments for the existence of dark matter in the Universe is
the existence of flat rotation curves in galactic disks. The existence of
such a flat rotation curves is accepted for several decades. As for some
references on this theme, we prefer to mention some of the standard textbooks
used in the world than an immense number of articles (the greatest part of which
the author has either way not in disposal); thus: Mihalas and McRae Routly
(1968), Vorontsov-Velyaminov (1978), Mihalas and Binney (1981), Kulikovskij
(1985), Scheffler and Els\H{a}sser (1988), Zeilik (1994).

\section{Radial Velocity Data and Trajectories}
Let us consider the following simple model (it can be an approximation
to real situation if we take into account that mass density exponentially
decreases with distance from the center of the Galaxy).

Let us suppose that the mass of the Galaxy is distributed in the form that
objects within a few kpc around the Sun move in Keplerian orbits.
(Data in Fernie and Hube (1968) show that stars in galactic disk move
in strongly noncircular orbits.)
We suppose that all objects of the galactic disk move in equatorial plane
of the Galaxy, for simplicity. Thus, we consider that a star and the Sun
move in galactic plane on Keplerian orbits characterized by the following sets
of quantities (their physical interpretation is standard):
$\{ p, e, \phi, \omega, R = p / [ 1 ~+~e ~ \cos
( \phi ~-~ \omega ) ] \}$ for the star, and
$\{ p_{o}, e_{o}, \phi_{o}, R_{o} = p_{o} / [ 1 ~+~e_{o} ~ \cos
\phi_{o}  ] \}$ for the Sun. Simple plane geometry yields
$\sin ( \phi ~-~ \phi_{o} ) = ( r~ \sin l ) / R$, where
$r$ is distance between the Sun and the star , $l$ is galactic
longitude of the star. Keplerian motion yields for radial velocity
of the star with respect to the Sun:
\begin{eqnarray}\label{1}
\frac{\Delta v_{r}}{\sqrt{\mu}} &=&
      \frac{1}{\sqrt{p}} ~ \left \{ \frac{R_{o}}{R}~ \sin l ~-~
       e ~ \left [ \sin \left ( \phi_{o} ~-~ \omega \right ) ~ \cos l ~-~
       \cos \left ( \phi_{o} ~-~ \omega \right ) ~ \sin l \right ] \right \}~-~
\nonumber \\
& & ~-~
      \frac{1}{\sqrt{p_{o}}} ~ \left \{ \left ( 1 ~+~ e_{o}~ \cos \phi_{o}
      \right ) ~ \sin l ~-~
      \left ( e_{o} ~ \sin \phi_{o} \right ) ~ \cos l \right \} ~,
\end{eqnarray}
where also
$R = \sqrt{R_{o}^{2} ~+~ r^{2} ~-~ 2~ r ~ R_{o} ~ \cos l}$ may be used.

Making Taylor expansion in $r / R_{o}$ (and to the order $e^{2}$) one can
receive
\begin{eqnarray}\label{2}
\frac{\Delta v_{r}}{\sqrt{\mu / R_{o}}} &=& \left [
	 1 ~-~ \sqrt{\frac{\mu}{p_{o}}} \left ( 1 ~+~ e_{o} \cos \phi_{o} \right )
	 / \sqrt{\frac{\mu}{R_{o}}} ~+~ \frac{e}{2} ~ \cos \alpha ~-~
	 \frac{e^{2}}{16} ~ \left ( 1 ~+~ \cos 2 \alpha \right ) \right ] ~
	 \sin l 
\nonumber \\
& & ~+~ \left [
	 \sqrt{\frac{\mu}{p_{o}}} ~ \left ( e_{o} \sin \phi_{o} \right ) ~
	 / \sqrt{\frac{\mu}{R_{o}}} ~-~ e ~ \sin \alpha ~+~
	 \frac{e^{2}}{4} ~ \sin 2 \alpha  \right ] ~
	 \cos l ~+~
\nonumber \\
& & ~+~ \frac{r}{R_{o}} ~ \left \{ \frac{1}{4}	\left [ 3 ~-~ \frac{e}{2} ~
     \cos \alpha ~-~
     \frac{e^{2}}{8} ~ \left ( 9 ~+~ \cos 2 \alpha \right ) \right ] ~
     \sin 2 l  ~-~ \right .
\nonumber \\
& &  \left . ~-~ \frac{e}{2} ~ \left ( \sin \alpha ~-~ \frac{e}{4} ~
     \sin 2 \alpha \right ) ~ \cos 2 l \right \}
~+~ O \left \{ \left ( r / R_{o} \right )
     ^{2} \right \} ~,
\end{eqnarray}
where $\alpha = \phi_{o} ~-~ \omega$. The quantity $\mu \equiv G ~ M_{GC}$
is given by mass of the galactic center and its surrounding; circular velocity
of the Sun would be $v_{o} = \sqrt{\mu / R_{o}}$.

In practice we make measurements of $\Delta v_{r}$ (Doppler effect) and we do not
know values of $e, \omega$ for individual stars. Thus, we make some assumptions.
At first, in order to obtain relations analogous to those which are
standardly used, we make averaging of Eq. (2) in $\omega$ -- supposed to be
independent on other parameters; thus, using, e. g.,
(it is supposed that orbits are randomly oriented)
$< \sin \omega > =< \cos \omega > = 0$,
$< \sin^{2} \omega > = < \cos^{2} \omega > =$ 1/2,
($< e^{k} ~ f ( \omega ) > ~=~ < e^{k} > ~ < f ( \omega ) >$)
-- all these relations correspond to the fact that we neglect
all sums containing odd powers of sines and cosines of $\omega$ --
 one receives
\begin{equation}\label{3}
\Delta v_{r} = \kappa_{1} ~ \cos l ~+~ \kappa_{2} ~ \sin l ~+~
		 A ~ r ~ \sin 2 l ~+~ r^{2} ~ \left \{
		 a_{1} ~ \sin l ~+~ a_{3} ~ \sin 3 l \right \} ~,
\end{equation}
where
\begin{eqnarray}\label{4}
\kappa_{1} &=& \sqrt{\mu / p_{o}} ~~e_{o} ~ \sin \phi_{o} ~, \nonumber \\
\kappa_{2} &=& \sqrt{\mu / R_{o}} ~~ ( 1 ~-~ ( 1 / 16 ) ~ < e^{2} > ) ~-~
 \sqrt{\mu / p_{o}} ~~( 1 ~+~ e_{o} ~ \cos \phi_{o} ) ~,
\nonumber \\
A &=& ( 3 / 4 ) ~\sqrt{\mu / R_{o}^{3}} ~~ ( 1 ~-~ ( 3 / 8 ) ~ < e^{2} > ) ~,
\nonumber \\
a_{1} &=& -~ ( 3 / 32 ) ~\sqrt{\mu / R_{o}^{5}} ~~ ( 1 ~-~ ( 9 / 4 ) ~ < e^{2} > ) ~,
\nonumber \\
a_{3} &=& +~ ( 3 / 32 ) ~\sqrt{\mu / R_{o}^{5}} ~~ ( 7 ~-~ ( 3 / 4 ) ~ < e^{2} > ) ~,
\end{eqnarray}
where $< e^{2} >$ denotes average value for large number of stars;
also higher order in $r / R_{o}$ was written in Eq. (3) in comparison
with Eq. (2).

\section{Solar velocity}

The important quantity in determining of rotation curve is the velocity
of the Sun -- velocity of the motion of the Sun around the galactic center.

Standard procedure corresponds to the following situation: We take into account
Eq. (3). Now, we will consider only terms with $\kappa-$s:
\begin{eqnarray}\label{5}
\kappa_{1} &=& \sqrt{\mu / R_{o}} ~( e_{o} ~ \sin \phi_{o} )
	       ( 1 ~-~ ( 1 / 2 ) ~e_{o} \cos \phi_{o} )  ~, \nonumber \\
\kappa_{2} &=& -~ \sqrt{\mu / R_{o}} \left \{ ~( e_{o} / 2 ) ~( \cos \phi_{o} ) ~
	       ( 1 ~-~ ( 1 / 4 ) ~e_{o} \cos \phi_{o} )  ~+~ ( 1 / 16 ) ~
	       < e^{2} > \right \} ~.
\end{eqnarray}
Now, the least square method reads:
\begin{equation}\label{6}
\sum \left \{ \Delta v_{ri} ~-~ \kappa_{1} ~ \cos l_{i} ~-~
	\kappa_{2} ~ \sin l_{i} \right \} ^{2} = min ~.
\end{equation}
Differentiation with respect to $v_{o} \equiv \sqrt{\mu / R_{o}}$ and putting
the result equal to zero, yields finally
\begin{equation}\label{7}
v_{o} = X / Y ~,
\end{equation}
where
\begin{eqnarray}\label{8}
X &=& ( e_{o} ~~ \sin \phi_{o} ) ~( 1 ~-~ ( 1 / 2 ) ~e_{o} \cos \phi_{o} ) ~
      \sum ( \Delta v_{ri}  ~ \cos l_{i} ) ~-~
\nonumber \\
  &-&  \left \{ ~( e_{o} / 2 ) ~( \cos \phi_{o} ) ~
	       ( 1 ~-~ ( 1 / 4 ) ~e_{o} \cos \phi_{o} )  ~+~ ( 1 / 16 ) ~
	       < e^{2} > \right \} ~
      \sum ( \Delta v_{ri}  ~ \sin l_{i} ) ~,
\end{eqnarray}
\begin{eqnarray}\label{9}
Y &=& ( e_{o} ~~ \sin \phi_{o} ) ^{2} ~ \sum \cos^{2} l_{i} ~+~
~(( 1 / 2 ) ~e_{o} \cos \phi_{o} ) ^{2} ~ \sum \sin^{2} l_{i} ~-~
\nonumber \\
  & & ~-~ ~e_{o} ^{2} ~ \sin \phi_{o} ~ \cos \phi_{o} ~
       \sum \sin l_{i} ~ \cos l_{i} ~.
\end{eqnarray}

However, correct procedure must use Eq. (2). Again, neglecting terms
proportional to $r / R_{o}$ we may write
\begin{eqnarray}\label{10}
\frac{\Delta v_{r}}{\sqrt{\mu / R_{o}}} &=& L_{1} ~ \sin l ~-~
       L_{2} ~ \cos l ~,
\nonumber \\
L_{1} &=& ( e  / 2 ) ~ \cos ( \phi_{o} ~-~ \omega ) ~ \{ 1 ~-~ ( e / 4 )
	 \cos ( \phi_{o} ~-~ \omega ) \}  ~-~
\nonumber \\
  & &	 ~-~   ( e_{o} / 2 ) ~ \cos \phi_{o}  ~
	       ( 1 ~-~ ( 1 / 4 ) ~e_{o} \cos \phi_{o} ) ~,
\nonumber \\
L_{2} &=& e ~ \sin ( \phi_{o} ~-~ \omega ) ~ \{ 1 ~-~ ( e / 2 )
	 \cos ( \phi_{o} ~-~ \omega ) \}  ~-~
\nonumber \\
  & &	 ~-~   e_{o} ~ \sin \phi_{o}  ~
	       ( 1 ~-~ ( 1 / 2 ) ~e_{o} \cos \phi_{o} ) ~.
\end{eqnarray}
Now, the least square method reads:
\begin{equation}\label{11}
\sum \left \{ \Delta v_{ri} ~-~ v_{o} ( L_{1i} ~ \sin l_{i} ~-~
	L_{2i} ~ \cos l_{i} ) \right \} ^{2} = min ~.
\end{equation}
Differentiation with respect to $v_{o} \equiv \sqrt{\mu / R_{o}}$ and putting
the result equal to zero, yields finally
($< \sin \omega > =< \cos \omega > = 0$,
$< \sin^{2} \omega > = < \cos^{2} \omega > =$ 1/2;
$< e^{2} >$ denotes average value for large number of stars)
the following approximation:
\begin{equation}\label{12}
v_{o} = X / Z ~,
\end{equation}
where
$X$ is given by Eq. (8) and $Z$ is given by the following equation:
\begin{eqnarray}\label{13}
Z &=& \{ ( 1 / 2 ) ~ < e^{2} > ~+~ ( e_{o} ~~ \sin \phi_{o} ) ^{2} \}
      ~ \sum \cos^{2} l_{i} ~+~
\nonumber \\
 & & ~+~ \{ ( 1 / 8 ) ~ < e^{2} > ~+~  (( 1 / 2 ) ~e_{o} \cos \phi_{o} ) ^{2} \}
      ~ \sum \sin^{2} l_{i} ~-~
\nonumber \\
  & & ~-~ ~e_{o} ^{2} ~ \sin \phi_{o} ~ \cos \phi_{o} ~
       \sum \sin l_{i} ~ \cos l_{i} ~.
\end{eqnarray}

Comparison of Eq. (7) and (12) yields that solar velocity must be in reality less
than it is supposed up to now.	( $< e^{2} > \approx e_{o}^{2}$ and standard
values of $\kappa-$s and $A$ yields that terms with $< e^{2} >$ in Eq. (13)
are not negligible (!) and one must expect that solar velocity is in several
tens of percent less than IAU standard.) Of course, rigorous procedure
for the model must use Eq. (2) with the least square method for parameters
$p_{o}$, $e_{o}$, $\phi_{o}$.

\section{Neutral Hydrogen and Solar Velocity}
Neutral hydrogen moves in galactic plane practically on circular orbits
(see Fig. 8-11, p. 491, Mihalas and Binney 1981). Thus, as a very good
approximation we can use for the measured radial velocity $\Delta v_{r}$
\begin{equation}\label{14}
\Delta v_{r} = ( \omega ~-~ \omega_{o} ) ~ R_{o} ~ \sin l ~,
\end{equation}
where $\omega \equiv \omega (R) = v(R) / R$ is angular velocity
(in this section) of the gas cloud in distance $R$ from the galactic center,
$\omega_{o} = v_{o} / R_{o}$ is the analogous quantity for the Sun (circular
orbit of radius $R_{o}$), $l$ is galactic longitude of the cloud
(see, e. g., Eq. (13) in Kla\v{c}ka 1997, or also arbitrary textbook
on galactic astronomy already mentioned).

Looking in a direction $l \in (0^{\circ}, 90^{\circ})$, $\Delta v_{r}$
(and $\omega$) reaches its maximum value for $R = R_{o} ~ \sin l$
(p. 471, Mihalas and Binney 1981). So, we can construct rotation curve
for the galactic disk:
\begin{equation}\label{15}
v(R) = \Delta v_{r} ~+~ v_{o} ~ \sin l ~, ~~~R = R_{o} ~ \sin l ~.
\end{equation}
Fig. 8-11 (p. 491, Mihalas and Binney 1981) yields, approximately,
\begin{equation}\label{16}
\Delta v_{r} ~ [km/s] = 180 ~-~ 2~ l~ [^{\circ}] ~,
\end{equation}
for $l \in (\approx 30^{\circ}, \approx 70^{\circ})$.

If we use IAU standard value $v_{o} =$ 220 km/s, Eqs. (15) and (16)
enable to construct rotation curve for our Galaxy. We receive standard flat
rotation curve in this way.

However, the consequence of the preceding section is that the solar velocity
$v_{o}$ must be less than 220 km/s. Is it possible to obtain an approximation
of real value of $v_{o}$ on the basis of neutral hydrogen only? It seems to us
that the answer is ``yes''. What is the physical interpretation of negative
values of $\Delta v_{r}$ for
$l \in (\approx 30^{\circ}, \approx 150^{\circ})$ in
Fig. 8-11 (p. 491, Mihalas and Binney 1981)?
Its minimum value can be approximated by relation
\begin{equation}\label{17}
\Delta v_{r} ~ [km/s] = -~ 110~ \sin l ~.
\end{equation}
Putting this into Eq. (14), one obtains
\begin{equation}\label{18}
v ( R ) ~ \frac{R_{o}}{R} = v_{o} ~-~ 110 ~ km/s ~.
\end{equation}
Eq. (18) suggests that we must admit that Eq. (17) physically corresponds to
the solar radial velocity -- radial velocity of a gas cloud which moves
with negligible velocity around the galactic center. Really, if we consider
that radius of the Galaxy in neutral hydrogen is analogous to that for other
spiral galaxies (see, e. g., Fig. 22, p. 88 in Vorontsov-Velyaminov 1978),
$R_{o} / R <$ 1/5 (approximately) and the left-hand side of Eq. (18)
is much less than 110 km/s. If we neglect the left-hand side of Eq. (18),
then we obtain
\begin{equation}\label{19}
v_{o} \approx ~110 ~km/s ~.
\end{equation}
Using this value (or, a little higher), we can again use
Eqs. (15) and (16) for constructing rotation curve for our Galaxy.
The result is that $v(R)$ is a decreasing function of $R$ even in the inner
part of the solar orbit ($R / R_{o} >$ 1/2, approximately)! If this is
true, another important consequence is that dark matter cannot exist
within the gas radius of the Galaxy ($\approx$ 50 kpc).

\subsection{A Short Comment}
This subsection concerns a short comment on the maximum radial
velocity along the line of sight at $l$.

The important physical relation is given by Eq. (14) and by the relation
$R = R_{o} ~ \sin l$. If we want to make Taylor expansion around the value
$R_{o}$, we have to make Taylor expansion of Eq. (14), and, finally,
we may put $R = R_{o} ~ \sin l$ into the obtained result.  Thus,
\begin{equation}\label{20}
\Delta v_{r} = 2 ~A ~R_{o} ~ \left ( 1 ~-~ \sin l \right ) ~ \sin l ~+~
    \frac{1}{2} ~ \omega_{o} '' ~
    R_{o}^{3} ~ \left ( 1 ~-~ \sin l \right ) ^{2} ~ \sin l ~+~ ...~,
\end{equation}
where $A = -~ \omega_{o} ' ~ R_{o}~ / 2$. Since $\omega = v ( R ) / R$,
it can be easily verified that
\begin{equation}\label{21}
\omega_{o} '' = v_{o} '' ~ /~ R_{o} ~+~ 4 ~A ~/ ~ R_{o}^{2}  ~.
\end{equation}
Putting Eq. (21) into Eq. (20), one finally obtains
\begin{eqnarray}\label{22}
\Delta v_{r} &=& 2 ~A ~R_{o} ~ \left ( 2 ~-~ \sin l \right ) ~
\left ( 1 ~-~ \sin l \right ) ~ \sin l ~+~
\nonumber \\
& & +~ \frac{1}{2} ~ v_{o} '' ~
    R_{o}^{2} ~ \left ( 1 ~-~ \sin l \right ) ^{2} ~ \sin l ~+~ ...~.
\end{eqnarray}
This result differs from Eq. (8-26) in
Mihalas and Binney (1981; p. 476): they are incorrect since they make
Taylor expansion of our Eq. (15) instead of Eq. (14).

\section{Rotation Curves of Other Galaxies}
``The real velocity of rotation equals $V = V_{r}$ ~ cosec $i$, where
$V_{r}$ -- observed velocity along the line of sight at a distance
$r$ along the observable long axis of a galaxy.'' (Vorontsov-Velyaminov 1978,
p. 85 -- translation from Russian)

The consequence of such a definition is that we must receive flat rotation
curves even for Keplerian motion! We will show it. We will consider
cosec $i =$ 1.

Keplerian motion of a star around the center of mass (a galaxy)
is described by the following equations:
\begin{eqnarray}\label{23}
\vec{v} &=& v_{R} ~ \vec{e_{R}} ~+~v_{T} ~ \vec{e_{T}}
\nonumber \\
v_{R} &=& \sqrt{\mu / p} ~~ e~ \sin ( \phi ~-~ \omega ) ~~,~
v_{T} = \sqrt{\mu / p} ~ [ 1 ~+~ e~ \cos ( \phi ~-~ \omega ) ]~.
\end{eqnarray}
The fixed coordinate system is defined by the relations
\begin{eqnarray}\label{24}
\vec{e_{R}} &=& +~\cos \phi ~ \vec{i} ~+~ \sin \phi ~ \vec{j} ~,
\nonumber \\
\vec{e_{T}} &=& -~\sin \phi ~ \vec{i} ~+~ \cos \phi ~ \vec{j} ~.
\end{eqnarray}
Eqs.(23)-(24) yield
\begin{eqnarray}\label{25}
\vec{v} \cdot \vec{i} &=& v_{R} ~\cos \phi ~-~v_{T} ~ \sin \phi ~,
\nonumber \\
\vec{v} \cdot \vec{j} &=& v_{R} ~\sin \phi ~+~v_{T} ~ \cos \phi ~.
\end{eqnarray}

As the observed velocity along the line of sight, for one object, we take
\begin{equation}\label{26}
\vec{v} \cdot \vec{j} = \sqrt{\mu / p} ~ ( \cos \phi ~+~ e~ \cos \omega ) ~,
\end{equation}
where Eq. (23) was used.

Now, we can put equation $p = R ~( 1 ~+~ e~ \cos ( \phi ~-~ \omega ) )$
into Eq. (26). Making averaging over $\omega$ and neglecting higher orders
of eccentricities, one finally receives
\begin{equation}\label{27}
< \vec{v} \cdot \vec{j} > _{\omega} = \sqrt{\mu / R} ~
				     ( 1 ~-~ e^{2} / 16 )~ \cos \phi ~.
\end{equation}
If we make observations at a distance $r$ along the observable
long axis of a galaxy, then $r = R~ \cos \phi$
(more correctly, there should be absolute value of $\cos \phi$), and,
$\cos \phi \in ( r / R_{G} , 1)$, where $R_{G}$ is radius of the galaxy.
Thus, Eq. (27) yields
\begin{equation}\label{28}
< \vec{v} \cdot \vec{j} > _{\omega} = \sqrt{\mu / r} ~
		  ( 1 ~-~ e^{2} / 16 )~ ( \cos \phi )^{3/2} ~.
\end{equation}
Since objects of various $\phi$ occurs
at a distance $r$ along the observable
long axis of a galaxy
($\cos \phi \in ( r / R_{G} , 1)$),
we take as $V_{r}$ the quantity
$ < ( < \vec{v} \cdot \vec{j} > _{\omega} ) > _{\phi}$:
\begin{equation}\label{29}
V_{r} = \sqrt{\frac{\mu}{r}} ~\left ( 1 ~-~ \frac{e^{2}}{16} \right )~
	\left \{ \sqrt{\frac{r}{R_{G}}} ~ \arccos \left ( 
        \frac{r}{R_{G}} \right ) \right \}^{-1} ~
	\int_{0}^{\arccos ( r / R_{G} )} ~( \cos \phi )^{3/2} ~d \phi ~.
\end{equation}
It can be easily verified that the function $V_{r} (r)$ represents
a flat rotation curve.



\section{Conclusions}

We have shown the difference between the standard procedures used in galactic
astronomy and the correct procedure for obtaining the velocity of the Sun
around the center of the Galaxy. The important result is that real velocity
must be less than the IAU standard. Calculations for real data are in preparation.

Independent method based on neutral hydrogen data for our Galaxy
confirms the result. Moreover, it yields value of about 110 km/s.
As a consequence,
the rotation curve for our Galaxy is not flat (!)
and the dark matter cannot exist
within the gas radius of the Galaxy ($\approx$ 50 kpc).

Calculations for other galaxies seems to be also consistent with decreasing
rotation curve.

We have shown that characterization of rotation curves as flat curves
may be the consequence of incorrect physical and mathematical
interpretation of the observed data. 
(The author would like to thank
to students for their patient listening of the author's lectures
on these themes (also on coordinate systems and transformations of
proper motions and on simple kinematics of galactic rotation)
during the last year.)

\end{document}